\documentclass{article}
\usepackage{ijcai24}

\usepackage{times}
\usepackage{soul}
\usepackage{url}
\usepackage[hidelinks]{hyperref}
\usepackage[utf8]{inputenc}
\usepackage[small]{caption}
\usepackage{graphicx}
\usepackage{amsmath}
\usepackage{amsthm}
\usepackage{booktabs}
\usepackage{algorithm}
\usepackage{algorithmic}
\usepackage{multirow}
\usepackage{makecell}
\usepackage{subcaption}
\usepackage[switch]{lineno}

\title{FairLENS: Assessing Fairness in Law Enforcement Speech Recognition}

\author{
Yicheng Wang$^1$
\and
Mark Cusick$^2$\and
Mohamed Laila$^2$\and
Kate Puech$^2$ \and
Zhengping Ji$^2$\and \\
Xia Hu$^3$\and
Michael Wilson$^2$\and
Noah Spitzer-Williams$^2$\and
Bryan Wheeler$^2$\And
Yasser Ibrahim$^2$
\affiliations
$^1$Texas A\&M University \
$^2$Axon Enterprise, Inc. \
$^3$Rice University\\
}

\begin{document}

\maketitle

\begin{abstract}
    Automatic speech recognition (ASR) techniques have become powerful tools, enhancing efficiency in law enforcement scenarios. To ensure fairness for demographic groups in different acoustic environments, ASR engines must be tested across a variety of speakers in realistic settings. However, describing the fairness discrepancies between models with confidence remains a challenge. Meanwhile, most public ASR datasets are insufficient to perform a satisfying fairness evaluation. To address the limitations, we built FairLENS - a systematic fairness evaluation framework. We propose a novel and adaptable evaluation method to examine the fairness disparity between different models. We also collected a fairness evaluation dataset covering multiple scenarios and demographic dimensions. Leveraging this framework, we conducted fairness assessments on 1 open-source and 11 commercially available state-of-the-art ASR models. Our results reveal that certain models exhibit more biases than others, serving as a fairness guideline for users to make informed choices when selecting ASR models for a given real-world scenario. We further explored model biases towards specific demographic groups and observed that shifts in the acoustic domain can lead to the emergence of new biases. 
\end{abstract}

\section{Introduction}
Automatic Speech Recognition (ASR) techniques facilitate tedious tasks such as speech-to-text transcription. In the Law Enforcement context, ASR models can assist officers by converting spoken language into written text, making it easier for them to review and search related material during complex investigations. These models can also be used to accelerate the creation of evidence transcripts, extracted from Police Body Worn Cameras, for courts of law, or detect situations where individual safety is at risk. Nevertheless, recent studies show that accuracy varies with demographic attributes~\cite{koenecke2020racial,mengesha2021don,rajan2022aequevox,dichristofano2023performance}. To ensure every voice is heard at the most critical moment, ASR systems used in law enforcement and justice must therefore be as resilient as possible to domain shifts and biases.

However, there is no standardized framework to assess and rank fairness levels for ASR models. While it might be straightforward to check an ASR model by observing its performances across different demographic groups~\cite{liutowards2022}, this leaves users with demographic group performance discrepancies and no principled methodology for determining the top-performing ASR model. Additionally, a comprehensive ASR fairness evaluation dataset including multiple, balanced demographic groups across law enforcement scenarios is also needed. Public ASR datasets only contain very limited demographic attribute labels \cite{librispeech} or unsatisfying data settings~\cite{liutowards2022,garofolo1993darpa}. 

To address the gaps, we present FairLENS, a \textbf{Fair}ness evaluation framework for \textbf{L}aw \textbf{EN}forcement \textbf{S}peech recognition. We first introduce an adaptable evaluation method that supports any combination or intersection of demographic groups, allowing for a fine-grained comparison of fairness levels among ASR models. We employ the Wilcoxon signed-rank test~\cite{wilcoxon1945individual} to examine the difference in paired data, helping us determine if significantly different fairness disparities exist between two models. 

To showcase the value of our fairness framework and identify potential ASR bias in the law enforcement domain, we collected a dataset that includes more than 2,300 unique native English speakers from a diverse set of attributes (i.e., Sex, Age, Race, Accent), resulting in over 300 hours of speech data simulating real-world law enforcement scenarios. All demographic attribute labels have been self-identified by the participants themselves. We have also made sure to cover various law enforcement contexts in different acoustic scenarios. Compared to other related datasets, our FairLENS dataset exhibits better coverage and a more balanced representation of various demographic groups.

Utilizing the FairLENS framework, we conducted a joint performance-fairness analysis, comparing 1 open-source and 11 commercial ASR models across diverse demographic groups and use cases. The results indicate that some models offer significantly fairer recognition accuracy compared to others. Our benchmark provides users with the insight to select ASR models that excel in both performance and fairness. 

Furthermore, we investigated the model biases towards specific groups. The results suggest that certain intersectional subgroups could be subject to more biases, potentially due to inherent model flaws or imbalanced training data. We also observed that changes in the acoustic domain (such as the introduction of noisy environments) cause model performance to degrade more for some groups compared to others.

Contributions presented in this paper can be summarized as follow:

\begin{enumerate}
    \item Building FairLENS, a comprehensive ASR fairness evaluation framework for law enforcement scenarios, including a general fairness evaluation method and a corresponding evaluation dataset.
    \item Assessing 12 state-of-the-art ASR models on our dataset and measuring fairness levels for all models.
    \item Highlighting that the biases towards specific demographic groups, combined with shifts in the acoustic domain, are critical concerns for ASR models.
\end{enumerate}

\section{Related Work}
\subsection{Fairness Evaluation on ASR Models}
Recent research has increasingly focused on identifying potential biases in ASR systems. It has been revealed that even the state-of-the-art ASR systems from Google, Amazon, and Apple, etc., show lower accuracy for African Americans~\cite{koenecke2020racial}, which causes substantial dissatisfaction~\cite{mengesha2021don}. DiChristofano et al.~\cite{dichristofano2023performance} also highlighted biases in ASR systems in relation to various accent groups. To assess ASR system fairness, Rajan et al. developed a testing framework AequeVox~\cite{rajan2022aequevox}, employing performance robustness on artificial noisy data as a fairness metric. However, this research had a limited scope, evaluating only three ASR models and relying on four public datasets that lacked comprehensive demographic analysis. Separately, Liu et al.~\cite{liutowards2022} leveraged modified Casual Conversations dataset~\cite{hazirbas2021towards} and four self-hosted ASR models to uncover performance gaps across different demographic groups. Despite these efforts, none of the studies provide a definitive fairness assessment or ranking among a wide range of state-of-the-art ASR models. 

\subsection{Datasets for Fairness Evaluation}
One of the first areas covered by fairness research was economics. For instance, the Adult dataset~\cite{kohavi1996scaling} for income prediction is labelled with sex, race, and age attributes. The German credit dataset~\cite{hofmann1994statlog}, used for credit score classification, is labelled with sex and race attributes as well. 

In computer vision, facial-related tasks are the most prone to fairness problems. CelebA~\cite{liu2015faceattributes} and RAF-DB~\cite{li2017reliable,li2019reliable} are two examples of datasets frequently used for fairness evaluation~\cite{xu2020investigating,wang2020towards}. However, these datasets have not been specifically designed for fairness evaluation purposes and lack a balanced representation of demographic groups, rendering them inadequate for fairness analyses across different subgroups.

Hazirbas et al.~\cite{hazirbas2021towards,hazirbas2022casual} presented two versions of the Casual Conversations video datasets for fairness evaluation on video tasks. The datasets include video recordings from thousands of participants and is annotated with age, gender, skin tone, and lighting. The first version was also modified to a transcription dataset~\cite{liutowards2022} for the ASR task. Yet, the original video dataset concentrates on both speech and facial information and was not designed with speech in mind, which makes the speech part a single form of indoor monologues. In this dataset, the data distribution is imbalanced across various groups, which affects the fairness of the ASR bias evaluation (Figure \ref{fig2b}). Hence, a comprehensive ASR fairness evaluation dataset is urgently needed.

\begin{table}[t]

\centering
\small
\begin{tabular}{|c|cccc|c|}
\hline
\multirow{2}{*}{\textbf{Model}}      & \multicolumn{4}{c|}{\textbf{Dataset}}   & \multirow{2}{*}{\textbf{Device}}                                      \\ \cline{2-5}
                               & \multicolumn{1}{c|}{Envir.}                                                                         & \multicolumn{1}{c|}{Spkrs} & \multicolumn{1}{c|}{Hours} & Cases &                                                                       \\ \hline
\multirow{3}{*}{\makecell[c]{\\ \\Trans.}} & \multicolumn{1}{c|}{\makecell[c]{Indoor\\Solo}}   & \multicolumn{1}{c|}{911}             & \multicolumn{1}{c|}{133}            & 4917       & {\makecell[c]{$95\%$ phone\\ $5\%$ AB3}}      \\ \cline{2-6} 
                               & \multicolumn{1}{c|}{\makecell[c]{Indoor\\Dialogue}}                                                                          & \multicolumn{1}{c|}{247}             & \multicolumn{1}{c|}{25}             & 333        & AB3                                                              \\ \cline{2-6} 
                               & \multicolumn{1}{c|}{\makecell[c]{Outdoor\\Dialogue}}                                                                         & \multicolumn{1}{c|}{246}             & \multicolumn{1}{c|}{68}             & 990        & AB3                                                              \\ \hline
Dict.                      & \multicolumn{1}{c|}{\makecell[c]{Indoor\\Solo}}  & \multicolumn{1}{c|}{905}             & \multicolumn{1}{c|}{96}             & 2391       & {\makecell[c]{$86.5\%$ phone\\ $13.5\%$ AB3}} \\ \hline
\end{tabular}

\caption{The data composition of the FairLENS dataset. Trans. is for Transcription ASR models and Dict. is for Dictation ASR models. The AB3 is Axon Body 3, a type of body-worn camera.}
\label{table1}
\end{table}

\section{FairLENS: A Fairness Evaluation Framework}
\label{sec:framework}

Fairness problems in ASR models have been uncovered by many recent studies, affecting many demographic attributes such as gender/sex~\cite{liutowards2022}, race~\cite{koenecke2020racial}, and accent~\cite{dichristofano2023performance}. However, there is no general evaluation framework for ASR tasks to benchmark fairness levels of models. Hence, we propose the FairLENS framework, especially targeting ASR models in law enforcement scenarios, but can be applied to other fairness evaluation tasks. This framework consists of two main parts: an adaptable fairness evaluation method and an ASR dataset.

\begin{table*}[!t]
\small
\centering
\begin{tabular}{|p{0.07\textwidth}|p{0.12\textwidth}|p{0.11\textwidth}|p{0.28\textwidth}|p{0.25\textwidth}|}
\hline
\textbf{Script} & \textbf{Sex}                                            & \textbf{Age}                                                                          & \textbf{Race}                                                                                             & \textbf{Accent}                                                                                                        \\ \hline
\begin{tabular}[c]{@{}l@{}} Solo \\ Speaker\end{tabular} & \begin{tabular}[c]{@{}l@{}}Female, \\ Male\end{tabular} & \begin{tabular}[c]{@{}l@{}}Adult ($19$-$55$),\\ Senior ($\ge56$),\\ Teen($\le18$)\end{tabular} & \begin{tabular}[c]{@{}l@{}} African American, \\ Asian, \\ Caucasian, \\Hispanic/Latino \end{tabular} & \begin{tabular}[c]{@{}l@{}} Midland, \\ New England/New York City, \\ Northern/North Central,\\ Southern, \\ Western \end{tabular} \\ \hline

\begin{tabular}[c]{@{}l@{}} Dialogue\end{tabular} & \begin{tabular}[c]{@{}l@{}}Female-Female, \\ Male-Male, \\ Female-Male\end{tabular} & \begin{tabular}[c]{@{}l@{}}Adult-Adult,\\ Senior-Senior,\\ Teen-Teen, \\ Other \end{tabular} & \begin{tabular}[c]{@{}l@{}} African American-African American, \\ Asian-Asian, \\Caucasian-Caucasian, \\Hispanic/Latino-Hispanic/Latino, \\ Other\end{tabular} & \begin{tabular}[c]{@{}l@{}} Southern-Southern, \\ Western-Western,\\ Other \end{tabular} \\
\hline

\end{tabular}
\caption{Demographic groups in the FairLENS dataset.}
\label{table2}
\end{table*}

\begin{figure*}[!t]
    \centering
    \includegraphics[width=\textwidth]{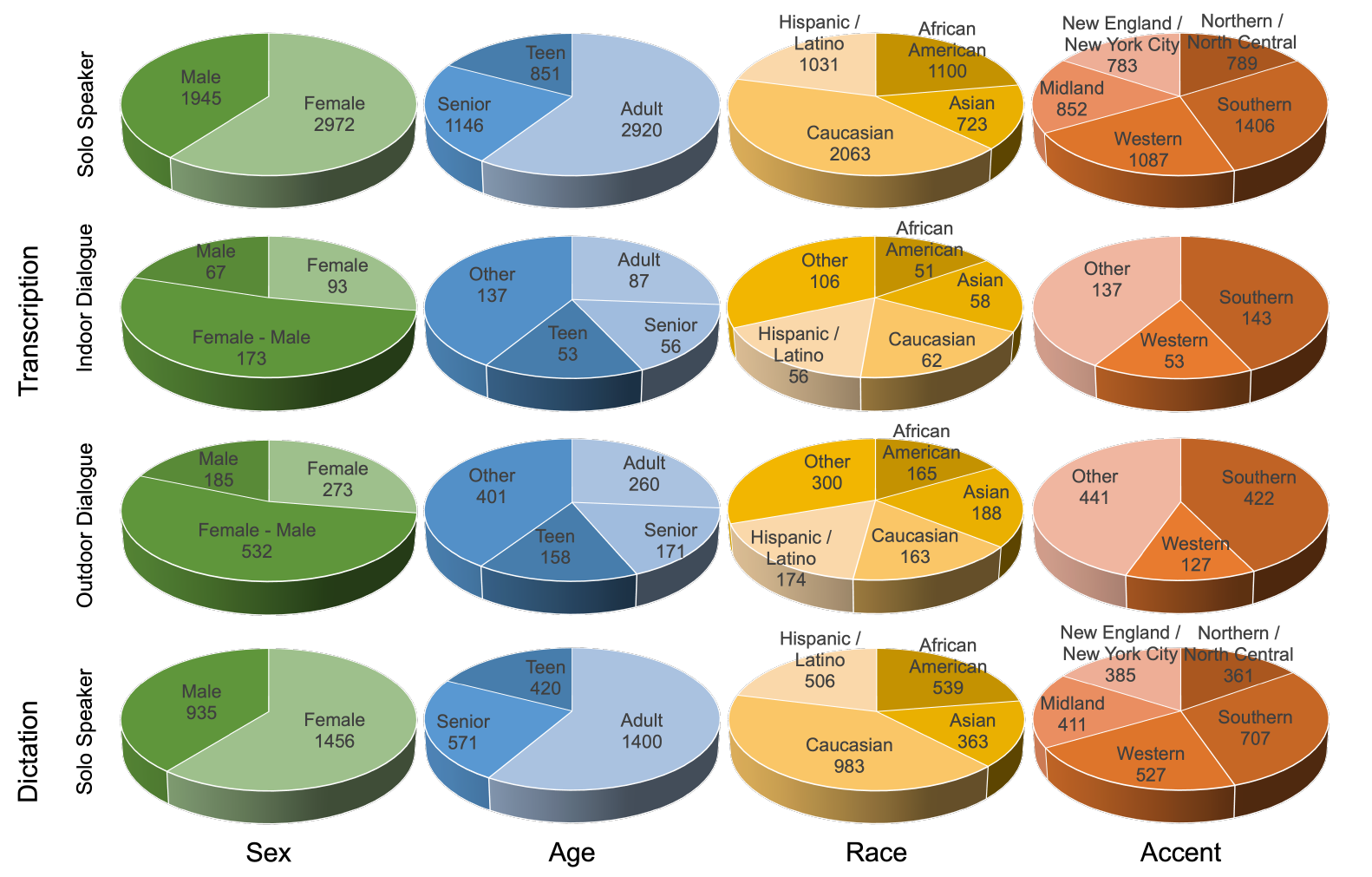}
    \caption{Data case distributions of the demographic groups in the FairLENS dataset.}
    \label{fig1}
\end{figure*}

\begin{figure*}[t]
     \centering
     \begin{subfigure}{\columnwidth}
         \centering
         \includegraphics[width=\columnwidth]{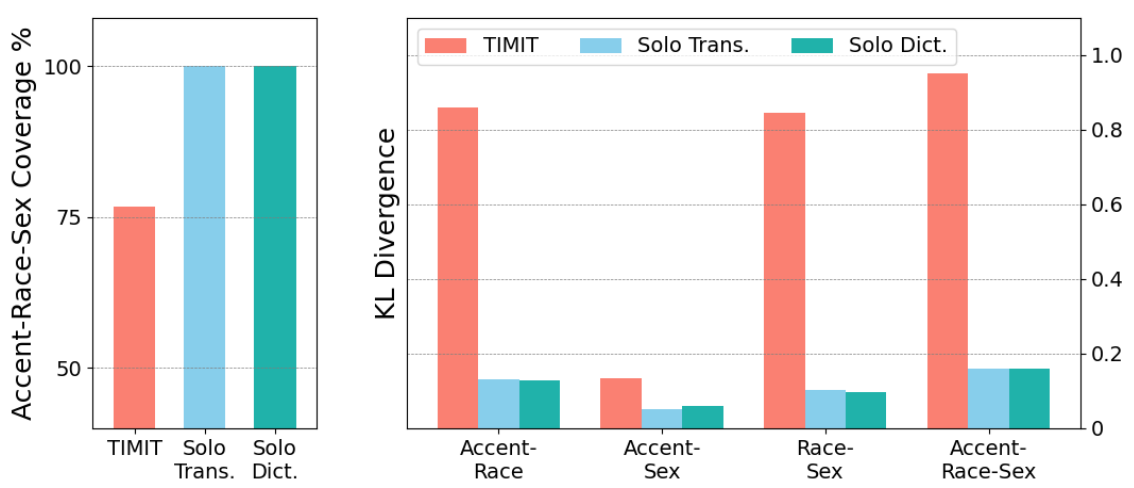}
         \caption{TIMIT v.s. FairLENS}
         \label{fig2a}
     \end{subfigure}
     \hfill
     \begin{subfigure}{\columnwidth}
         \centering
         \includegraphics[width=\columnwidth]{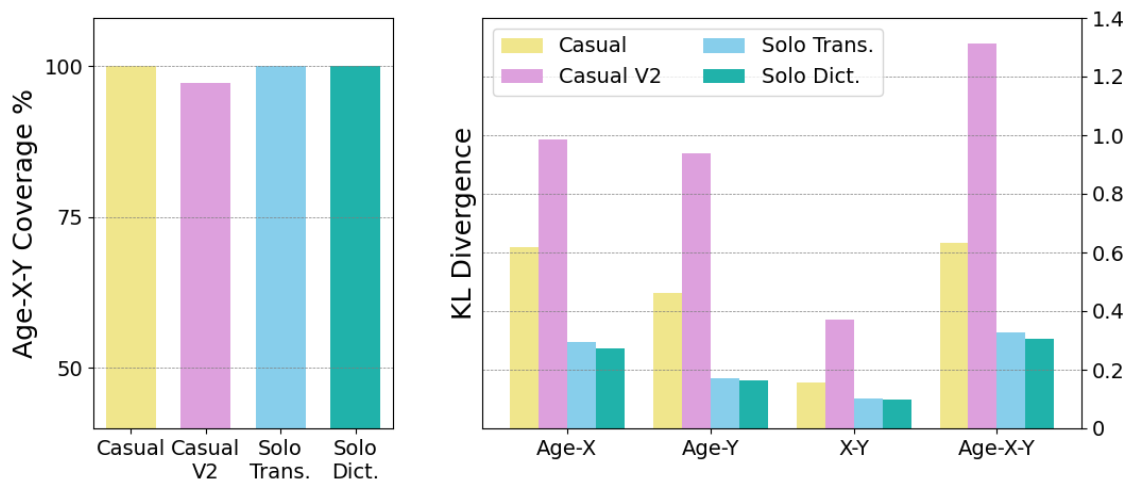}
         \caption{Casual Conversation v.s. FairLENS.}
         \label{fig2b}
     \end{subfigure}
        \caption{Comparing the FairLENS dataset with others.}
        \label{fig2}
\end{figure*}

\subsection{FairLENS Evaluation Method}

\textbf{ASR Performance Metric.} Word error rate (WER) is the common metric~\cite{librispeech,baevski2020wav2vec,liutowards2022} to measure the performances of ASR models. The bias-corrected and accelerated bootstrap method (BCa)~\cite{efron1986bootstrap} is employed to generate confidence intervals with $p<0.05$. If two models have overlapping confidence intervals, the performance difference between the two models is not significant.

\noindent \textbf{Fairness Evaluation Method.} The main challenge for this task is to synthesize multiple performance disparities between demographic groups to compare the fairness of the models overall. A straightforward approach is to rank the models on each of the groups and calculate the average rankings. Yet, it is possible that two models have no significant differences in fairness level, which should be detected via a statistical test. The mean-ranks post-hoc~\cite{mcdonald1967rank,nemenyi1963distribution} test is the simplest method to address this issue. While applying the test to two models, the result is related to the number of other models, which is contradictory. Instead of the mean-rank test, Benavoli et al.~\cite{benavoli16a} proposed to use the Wilcoxon signed-rank test~\cite{wilcoxon1945individual} from which the results are irrelevant to other models, which meets the requirements of the fairness evaluation task. 

Here we define our fairness evaluation metric: average WER disparity. This metric measures the mean WER distance to the overall average WER of a model. Suppose that there are $K$ types of subgroups, we can get $K$ average WER results for each subgroup: $\text{WER}_{k}, k=1, 2, ..., K$. 

Then we take the mean WER on the whole dataset denoted by $\overline{\text{WER}}$ as the baseline and compute the disparity (absolute distance) between $\text{WER}_{k}$ and $\overline{\text{WER}}$: 
\begin{equation}
    d_{k} = |\text{WER}_{k}-\overline{\text{WER}}|,
\end{equation}
so that we can acquire a series of disparities $\{d_{k}\}$ for each model. 

To determine the model's overall fairness, we can calculate the average WER disparity: 
\begin{equation}
   \Bar{d} = \sum_{k=1}^K d_{k}/K.  
\end{equation}

However, this average WER disparity $\Bar{d}$ doesn't show the significance of the fairness differences between any two models. Therefore, we conduct the Wilcoxon signed-rank test (W-Test) on the paired disparities for any two models:
\begin{equation}
    \text{W-Test}(\{d_{k}^{\text{model}_1}\},\{d_{k}^{\text{model}_2}\})
\end{equation}
to assess the significance of the fairness differences between the two models, i.e. if $p<0.05$. This statistical method helps determine if the one tested model is fairer than the other one with high confidence. The details of conducting the W-Test are shown in Appendix.
 
As a comparison, the authors in another ASR fairness study~\cite{liutowards2022} only calculate the relative WER gaps of different groups to illustrate the fairness levels, which lacks of the estimation of the uncertainty.

While our proposed fairness evaluation method is tailored for ASR tasks, it is a general approach that can be adapted to any other fairness evaluation tasks on corresponding datasets with multiple and balanced demographic groups. Users can easily replace the WER with any other performance metric to match their specific tasks. For instance, in a facial recognition task, one can leverage the recognition accuracy as the base metric then calculate the accuracy disparities for different groups as well as the average accuracy disparity, then conduct the Wilcoxon signed-rank test to examine the significance of the fairness. An example can be found in Appendix.

\subsection{FairLENS Dataset}
\noindent\textbf{Data Collection.} In constructing the FairLENS dataset, we followed three main principles:

\begin{enumerate}
    \item \textbf{Demographic Diversity}: A robust fairness evaluation dataset should encompass a board demographic range.
    \item \textbf{Balanced Representation}: Beyond mere demographic group balances, a fairness dataset should also achieve balance within intersectional subgroups. For instance, while ensuring a balance between Female/Male and Asian/Caucasian categories is essential, it is equally important to have similar numbers of data from Asian females and Caucasian females.
    \item \textbf{Self-Identification}: Drawing from prior studies~\cite{hazirbas2021towards,hazirbas2022casual}, we ensured that participants self-identify their sensitive attributes. This method not only reduce annotation costs but also avoid biases that might arise from external judgments.
\end{enumerate}


We conducted multiple rounds of data collection while monitoring the amount of data for each demographic subgroup, and intentionally collected data for under-represented demographic groups and subgroups. In practice, we executed two data collection procedures in three acoustic domains. First, we remotely collected data using a mobile app. Participants read provided scripts to their mobile devices in a room with minimal background noise, which allowed us to source many diverse participants. Second, we collected data onsite using body-worn cameras (BWC) to record scripted dialogues, in both quiet indoor and noisy outdoor settings. 

Four main categories of scripts were created: scripted and freeform monologues, scripted dialogues, and scripted dictations (police report dictation). The scripts are either totally scripted, e.g. Miranda Rights, or semi-scripted leaving some freeform entities and descriptions to the participants. A subset of monologue scripts and dictation scripts took inspiration from officer incident reports. The dialogue scripts took inspiration from real-world officer-civilian interactions. The freeform discussion prompts were created with the objective to tease out more accent- or locale-specific terminology with regard to common incidents that may be described to an officer: such as a car accident recounting, or a witness description of a potential suspect. Example scripts are provided in Appendix.

\begin{figure*}[t]
     \centering
     \begin{subfigure}{0.49\textwidth}
         \centering
         \includegraphics[width=0.97\columnwidth]{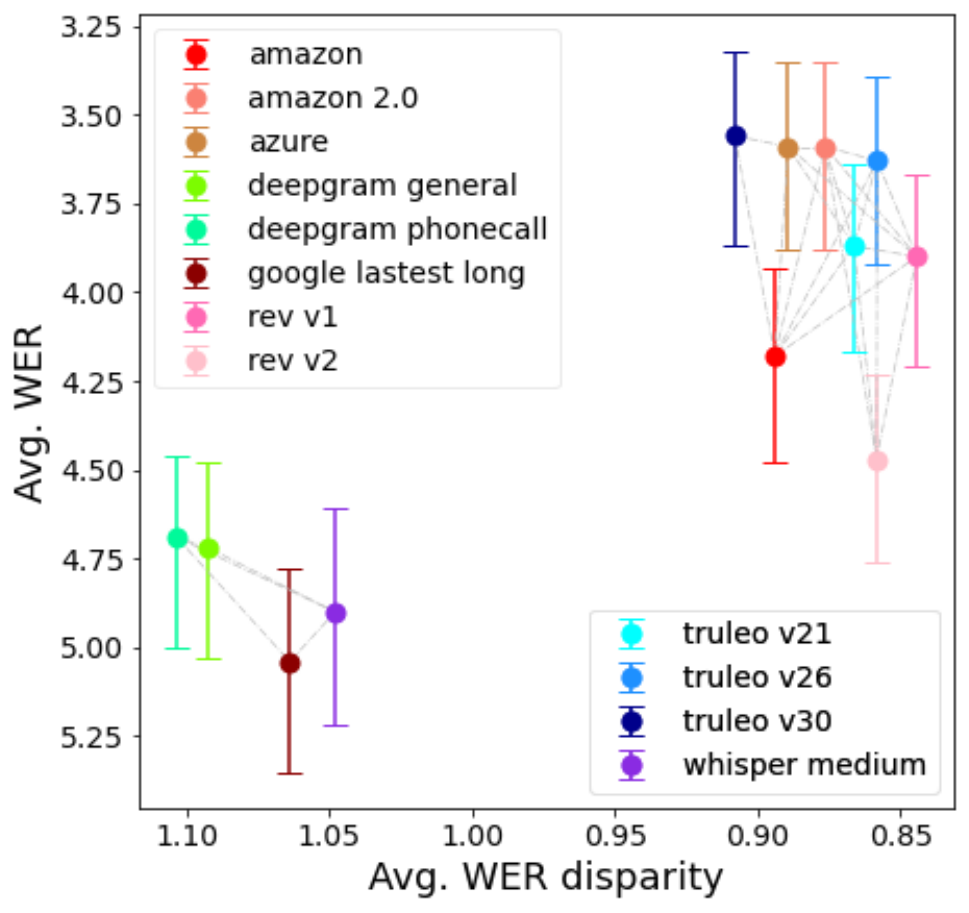}
         \caption{Solo speaker transcription dataset.}
         \label{fig3a}
     \end{subfigure}
     \hfill
     \begin{subfigure}{0.49\textwidth}
         \centering
         \includegraphics[width=0.97\columnwidth]{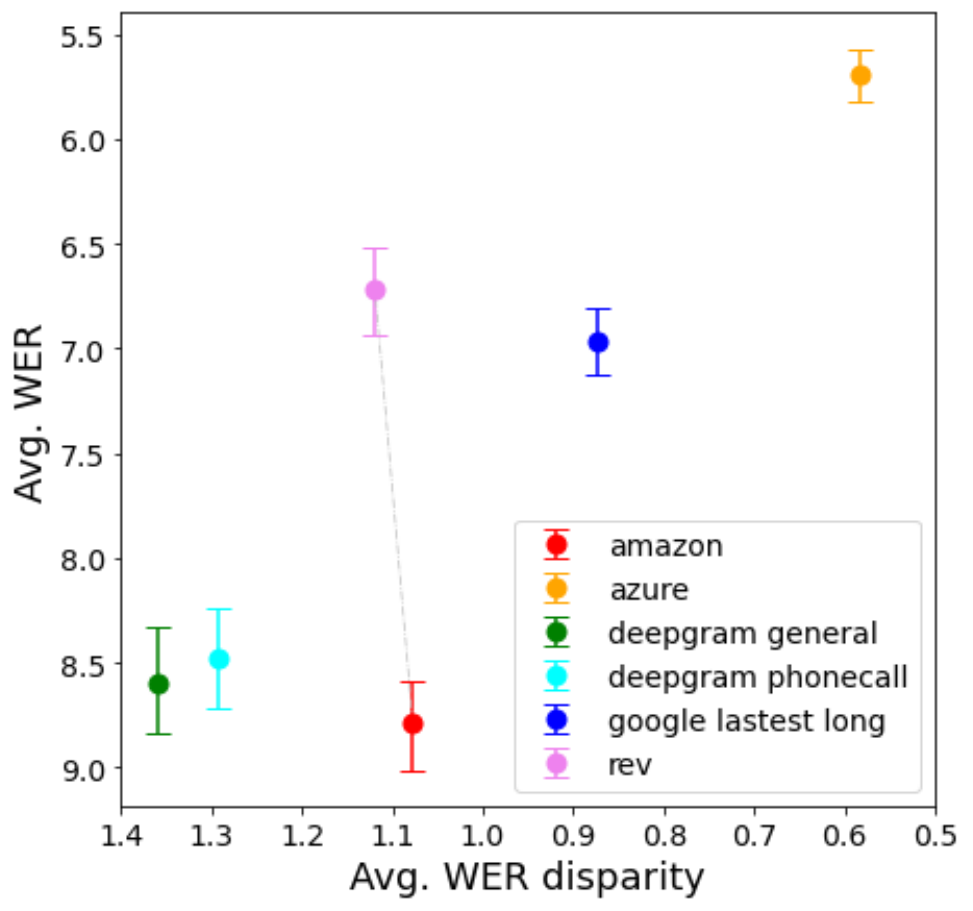}
         \caption{Solo speaker dictation dataset.}
         \label{fig3b}
     \end{subfigure}
         \caption{Joint performance-fairness evaluation results.}
        \label{fig3}
    \vspace{-0.5cm}
\end{figure*}

\begin{figure*}[t]
    \centering
    \includegraphics[width=\textwidth]{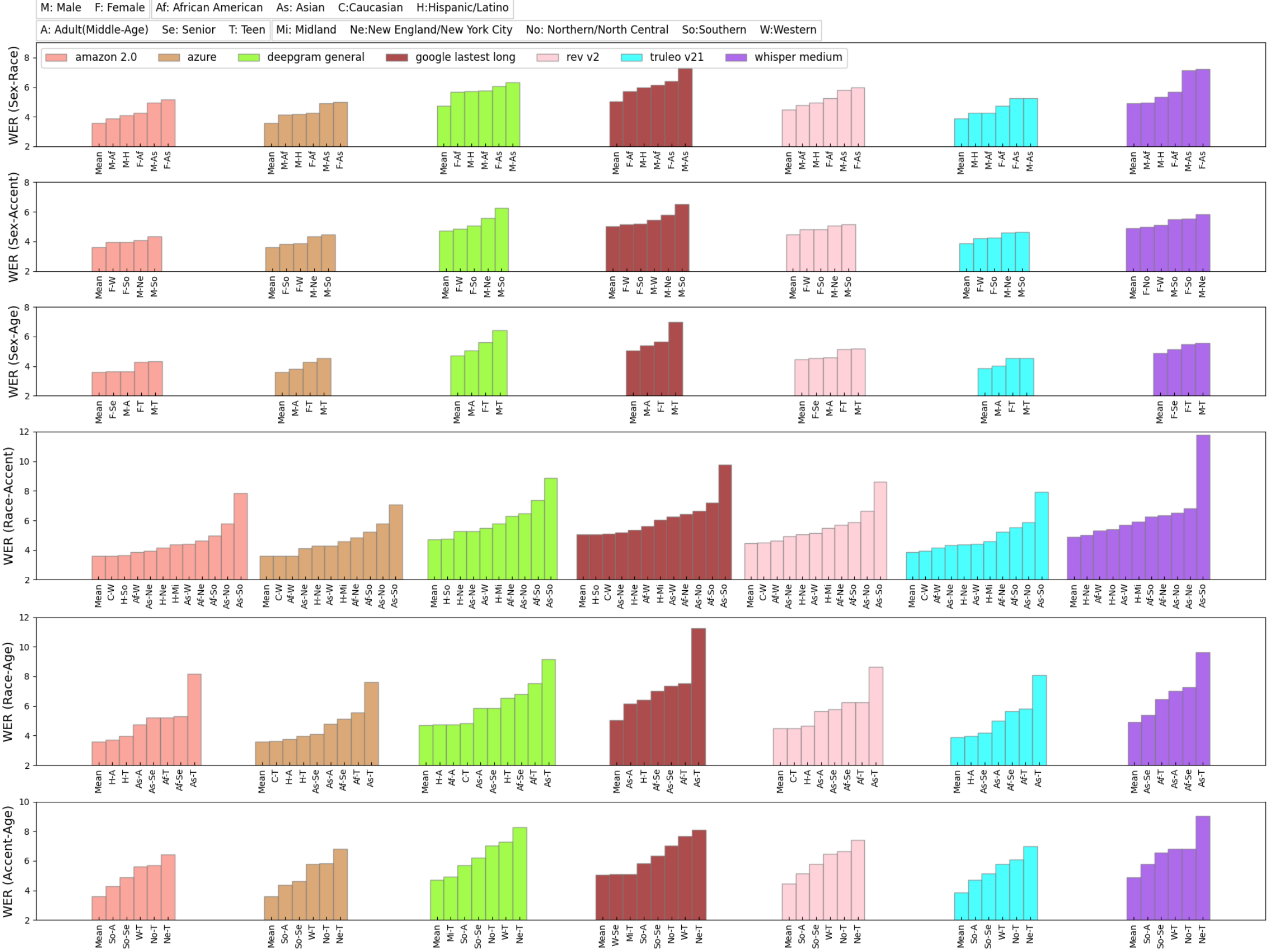}
    \caption{The performance of ASR models across different demographic subgroups on the solo speaker transcription dataset. We only shows those results that are worse than the average WER.}
    \label{fig4}
\end{figure*}

\begin{figure*}[t]
    \centering
    \includegraphics[width=\textwidth]{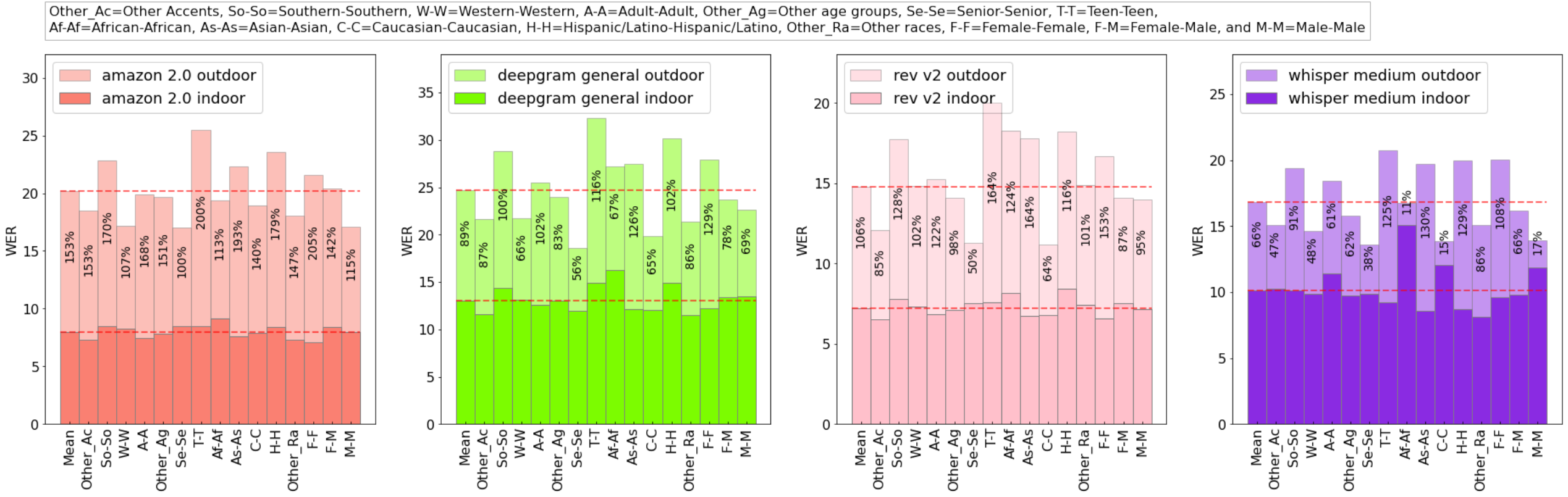}
    \caption{The performance of ASR models across different demographic groups on the indoor and outdoor dialogue datasets. The two dashed red lines in each graph represent the mean WER of the model on the indoor and outdoor dialogue datasets, respectively. The WER variations are shown on the bars.}
    \label{fig5}
\end{figure*}

\noindent\textbf{Data Composition.} 
As shown in Table \ref{table1}, FairLENS dataset is composed of four independent datasets for two types of models: transcription and dictation. The transcription ASR models convert a batch of existing audio recordings to texts while the dictation models stream audio-to-text conversion. In total, there are more than $300$-hours of audio data and more than $2,300$ participants in the FairLENS dataset. 

Our FairLENS dataset provides abundant self-identified sensitive labels shown in Table \ref{table2}: Sex, Age, Race, and Accent.  

\textbf{Sex}: The sex label covers two biological sexes, Male and Female \cite{census_gov_2021}. Due to the differences in pronunciation habits and average voice frequencies, ASR models might have varying recognition accuracy for males and females. Relevant research has also shown the performance gap of different genders~\cite{liutowards2022}. 

\textbf{Age}: For the age groups, we separated the ages into three buckets: Adult, Senior, and Teen. We created these categories based on the characteristics of the human voice, which changes with age. Two classic acoustic metrics: average fundamental frequency (F0) and signal-to-noise ratio (SNR), are utilized to illustrate: according to the results in a human acoustic survey~\cite{stathopoulos2011changes}, the F0 value of human voices declines rapidly before adulthood, then becomes almost steady, and slowly increases after the age of $50$ to $60$, while the SNR value increases before adulthood, then steady at the middle age, and decreases after $50$. Therefore, we split the age into three parts using $18$ and $55$ years as the cut-off. 

\textbf{Race}: We choose four main races: African American, Asian, Caucasian, and Hispanic/Latino, which together account for about $95\%$ of the total population in the U.S.~\cite{jin_talbot_wang_2021}. Recent studies~\cite{koenecke2020racial,mengesha2021don} have revealed the fairness problem of ASR models on different races, especially towards the minority groups.

\textbf{Accent}: The accent attribute has $5$ possible labels New England/New York City, Northern/Northern Central, Southern, Western, Midland, based on the categories of the U.S. English dialects~\cite{Potter2023-rr}. There is also evidence of accent bias in ASR models~\cite{dichristofano2023performance}.

Specifically for the two dialogue datasets, since there are two speakers in one audio clip, a demographic label for a dialogue is the combinations of the two original labels. For example, the sex attribute can take tow possible values, Male and Female, but for the dialogue datasets, the sex label of an audio clip is Female-Female, Male-Male, and Female-Male. The dialogue datasets were collected in a more realistic onsite setting, which resulted in insufficient data for some demographic groups. To follow our principles, these groups were categorized as 'Other', such as Senior-Teen in the age attribute, and Western-Midland in the accent attribute. 



The distributions of demographic groups of each dataset are shown in Figure \ref{fig2}. The solo speaker transcription and the solo speaker dictation datasets have enough data for all intersectional subgroups (combinations of two demographic attributes, e.g., sex-age subgroups).  In this case, we are able to conduct a fine-grained model fairness analysis. On the two dialogue datasets (respectively indoor and outdoor), we can observe the impact of background noise on the ASR models' performance (accuracy and bias).

\noindent\textbf{Comparison to Existing Datasets.} We use two metrics to quantitatively evaluate the ethical quality of ASR datasets. The first metric is Intersectional Coverage, which is defined as the percentage of subgroups with sufficient data coverage amongst all possible intersectional subgroups. A higher coverage means that a dataset includes more demographic subgroups, and hence is of better quality. Another metric for the attribute label balance evaluation is the KL Divergence between the data distribution of subgroups and a uniform distribution. A well-balanced dataset should have a small KL Divergence, such that there are similar numbers of data for every subgroup for fair comparisons.

We choose TIMIT \cite{garofolo1993darpa} and Casual Conversations~\cite{hazirbas2021towards,hazirbas2022casual} as baselines to compare against our Solo Transcription and Solo Dictation datasets. There are no accurate age labels in TIMIT so we only compared the Accent-Race-Sex attributes. In Casual Conversations, there are no sex and race attributes but instead related gender and skin tone labels. Therefore we use Age-X-Y labels, where X represents the Race/Skin Tone attribute, and Y is the Sex/Gender attribute for FairLENS/Casual Conversations, respectively. The detailed settings are in Appendix. 

The comparison results are shown in Figure \ref{fig2}. Our FairLENS dataset exhibits the same or higher Intersectional Coverage ($100\%$) and a much lower KL Divergence, which demonstrates that the data distribution of our dataset is more balanced than others.

\section{Fairness Evaluation Results}
\subsection{Joint Performance-Fairness Evaluation}
Using the framework above, we are able to assign the measured models to the two-dimensional performance-fairness evaluation plane. With the confidence interval and the results of the Wilcoxon signed-rank test, we can compare and choose the models with better average performance and higher fairness levels. $11$ commercial engines, including models from \textit{Amazon}~\cite{AmazonASR}, \textit{Azure}~\cite{AzureASR}, \textit{Deepgram}~\cite{DeepgramASR}, \textit{Google}~\cite{GoogleASR}, \textit{Rev}~\cite{RevASR}, and \textit{Truleo}~\cite{TruleoASR}, and one open-source \textit{Whisper}~\cite{WhisperASR} medium model, are evaluated on the solo transcription dataset. $6$ models that support streaming conversion are evaluated on the solo dictation dataset. The results are shown in Figure \ref{fig3}. Models closer to the upper right have better performance and fairness. If the error bars of the data points corresponding to the two models overlap, it indicates that there is no significant difference in their average performance (WER). Meanwhile, we connect two models with dashed gray lines if they do not have significant ($p\ge 0.05$) fairness disparity according to the Wilcoxon signed-rank test. 

In Figure \ref{fig3a}, we observe two fairness clusters of ASR transcription models. The models on the top right clusters are significantly ($p<0.05$) fairer than those on the bottom left. The Rev v1~\cite{RevASR} is recognized as the fairest among tested ASR models but the superiority is not significant. The Truleo V30~\cite{TruleoASR}, attains the lowest mean WER, yet its performance is not markedly superior to close contenders like Amazon 2.0 and Azure models.

Figure \ref{fig3b} illustrates the results for the solo speaker dictation dataset. It is evident that the Azure~\cite{AzureASR} model demonstrates a significantly lower average WER and WER disparity compared to its competitors. More results on the indoor and outdoor dialogue datasets are shown in Appendix.

An additional observation is that models belonging to the same series usually exhibit comparable levels of performance and fairness. This similarity suggests a likelihood of shared model structures and the use of identical training datasets among these models.

These results accurately indicate the average performance and fairness levels of the models, as well as the significance of the differences. In essence, we provide a comprehensive ranking of the ASR engines, offering users a clear reference for ASR model selection.  

\subsection{Biases toward Specific Demographic Groups}

Identifying specific demographic groups for whom current ASR models exhibit bias is also a main intent behind the framework. Therefore, we investigated the performance of different ASR models across all types of subgroups. The results on the solo speaker transcription dataset are shown in Figure \ref{fig4}. The performance of the ASR models is notably worse for certain demographic groups/subgroups, and the groups/subgroups affected are essentially the same across different models. Sex appears to be the least impacted attribute, while there are clear biases towards Asian, African American, Teens, and Southern Accent, as well as the subgroups formed by their combinations. 

Most of the models tested are not open-source, so we can only hypothesize about the reasons for this common fairness issue. First, since the state-of-the-art ASR models are based on some similar structures like multi-head attention modules, the models have exhibited biases during the training process when analyzing features specific to certain demographic groups, thereby impacting their final performance. Second, the imbalanced training datasets used by these models may also lead to biased results. The conclusion is clear: more effort in the ASR fairness domain is needed for ASR engines to be used with little human supervision in the context of Law Enforcement.

\subsection{Acoustic Domain Shift}
By leveraging the two dialogue datasets, we observed another fairness issue in different acoustic environments, as shown in Figure \ref{fig5}. As the background noise increases, the average performance of the model deteriorates. However, at the same time, we observed that the degree of performance degradation in certain groups is significantly more pronounced, e.g., Teen-Teen, Southern-Southern, Asian-Asian, and Hispanic-Hispanic. 

This disparity in performance degradation across different groups likely arises from a lack of diverse and representative training data, particularly in terms of varying accents, speech patterns, and noise conditions, leading to the model's diminished ability to accurately process and recognize speech from these specific demographics under noisy environments. This issue highlights the importance for ASR model providers to augment their training datasets to cover different acoustic conditions.

\section{Conclusion \& Future Work}

In this paper, we present FairLENS, an adaptable fairness evaluation methodology. Utilizing a dataset that is both demographically and acoustically diverse from the law enforcement field, we assess the performance and fairness of several ASR models, paying special attention to the significance. Our observations also highlight biases in ASR systems towards certain groups and issues related to acoustic domain shifts. These problems likely stem from a lack of diversity in training data and inherent issues within the model structures themselves. In the future, we aim to enhance and broaden our dataset by incorporating a wider range of demographics and script categories, e.g., non-native speakers and freeform dialogue. Our goal is to provide the community with a principled approach to perform comprehensive fairness assessments. We believe that with diverse data and principled fairness measurements, every voice can be equally heard.

\appendix

\section{Wilcoxon Signed-Rank Test}

\textbf{Wilcoxon signed-rank test} \cite{wilcoxon1945individual} is a non-parametric hypothesis testing which is utilized to detect if the medians of two paired samples $X_i$ and $Y_i$ for any $i$ are significantly different, i.e. the differences $Z_i=X_i-Y_i$, where $i=1, ..., M$, are symmetric about 0 \cite{conover1999practical}. The null hypothesis and alternative hypothesis are:

\textbf{Null Hypothesis $H_0$}: The observations $Z_i=X_i-Y_i$ are symmetric about 0.

\textbf{Alternative Hypothesis $H_1$}: The observations $Z_i=X_i-Y_i$ are not symmetric about 0.

Sort $|Z_1|, |Z_2|, ..., |Z_M|$ from small to large. Let $R_i$ be the rank of $|Z_i|, i=1, ..., M$.

The positive- and negative-rank
 sum are defined as the sum of the positive and negative signed rank, respectively:
 
\begin{align*}
   T^+ &= \sum_{1\le i \le M, Z_i>0}{R_i} \\
   T^- &= \sum_{1\le i \le M, Z_i<0}{R_i}
\end{align*}

Trivial proof: $T^++T^-=\frac{M(M+1)}{2}$. Therefore, we can use either $T^+$ or $T^-$ be the test statistic. It is also easy to show that the means and variances of $T^+$ and $T^-$ under the null hypothesis are \cite{pratt2012concepts}:

\begin{align*}
    E[T^+] &= E[T^-] = \frac{M(M+1)}{4}\\
    Var[T^+] &= Var[T^-] = \frac{M(M+1)(2M+1)}{24}
\end{align*}

Then the $T^+$ or $T^-$ can be normalized with the mean and variance to approximates normal distribution. The z-score is defined as

\[
z = \frac{T^+ \text{(or } T^- \text{)} -E[T^+]}{\sqrt{Var[T^+]}} = \frac{T^+ \text{(or } T^- \text{)} - \frac{M(M+1)}{4}}{\sqrt{\frac{M(M+1)(2M+1)}{24}}}.
\]

In some cases, if there are zero ties ($Z_i=0$) or non-zero ties ($R_i = R_j$), the mean and the variance need tie corrections \cite{cureton1967normal}:
\begin{align*}
    E[T^+] &= \frac{M(M+1)}{4} -\frac{o(o+1)}{4},\\
    Var[T^+] &= \frac{M(M+1)(2M+1)}{24} - \frac{o(o+1)(2o+1)}{24}\\
    &-\frac{\sum_{j}{t_j^3-t_j}}{48},
\end{align*} where $o$ is the number of all zero $Z_i$, and $t_j$ is the number of tied observations of each non-zero tied group.

If the z-score is large or small such that the corresponding significance level (p-value) exceeds threshold $\alpha$ (usually $\alpha = 0.05$), we can reject the null hypothesis and accept the alternative one, i.e., the paired samples do have significant different medians.

\section{Adaptation to Another Fairness Evaluation Task}

Our method using the Wilcoxon signed-rank test (W-Test) provides a statistically robust way to measure fairness level between different ASR models. Models showing lesser average disparity across groups are deemed fairer. Importantly, the W-Test is domain-agnostic, and can be generalized to evaluate disparities of different domains. To illustrate this, we present one more example on the face recognition fairness task~\cite{Yu_2021_ICCV}. This application reinforces the utility of the W-Test as a generalizable tool for fairness assessment across diverse computational domains, highlighting its potential for widespread adoption in fairness evaluations.

In this example, we compare the face recognition accuracy of a baseline model (A) and a proposed model (B) across different demographic groups: [African, Caucasian, South Asian, East Asian]. The face recognition accuracy results of the two models are:
\begin{align*}
    \text{Acc}(A) &= [89.5, 94.3, 93.4, 72.5], \\
    \text{Acc}(B) &= [91.2, 92.3, 91.7, 78.0].
\end{align*}

Then we can get the accuracy disparity results:

\begin{align*}
d(A) &= |\text{Acc}(A)-\overline{\text{Acc}(A)}| =  [2.075, 6.875, 5.975, 14.925],\\
d(B) &= |\text{Acc}(B)-\overline{\text{Acc}(B)}| =  [2.900, 4.000, 3.400, 10.300].
\end{align*}

The Average Accuracy Disparities of model A and B are: 
$\overline{d(A)}=7.46 > \overline{d(B)}=5.15$, 
which indicates that B is fairer than A on facial recognition. 

Then, we implement the Wilcoxon Signed-Rank Test $\text{W-Test}(d(A), d(B))$ and get p-value = 0.25, which is larger than the threshold $\alpha=0.05$. This p-value means that model B is not significantly fairer than A.

\section{Exemplar Scripts} \label{apdx_b}

\subsection{Solo Example}

My name's Pat Randall. How you doing, sir? 

I saw an unattended bag, and I was going to present it to the gate agent when he was done boarding. Because he was unaware. It’s not like I found an unattended bag and didn't know what was in it when I was bringing it onboard the airplane. 

I did check the bag. It was secure. But I wasn't able to talk to the gate agent. As far as about the bag. I still had to do my own checks and stuff like that. I wasn't going to do anything with the bag or anything like that. I was making sure that it, that it was all right.

But I did do the check before. I did my checks and I also checked the bag. It was secure. But I was waiting on the gate agent to make sure that, um, I spoke to them. But I wasn't trying to do anything with it or anything like that.

Oh, yeah. Here’s my driver’s license and ID. I'm from Florida. Jacksonville.

Sir, as I was walking to the gate area, I saw an unattended bag. This bag right here. I was just waiting until we was all secure and stuff like that and give it right back to the gate agent. But it was secured. I checked it.

There's nothing inside that's harmful or anything like that. I wasn't going to do anything with it except give it to the gate agent so she could take care of it from there.

It was near the bathroom, sir, and I took it on the plane with me. My intentions were good.

\begin{figure*}[!h]

\centering
\rotatebox{90}{\begin{minipage}{0.96\textheight}
\includegraphics[width=\textwidth]{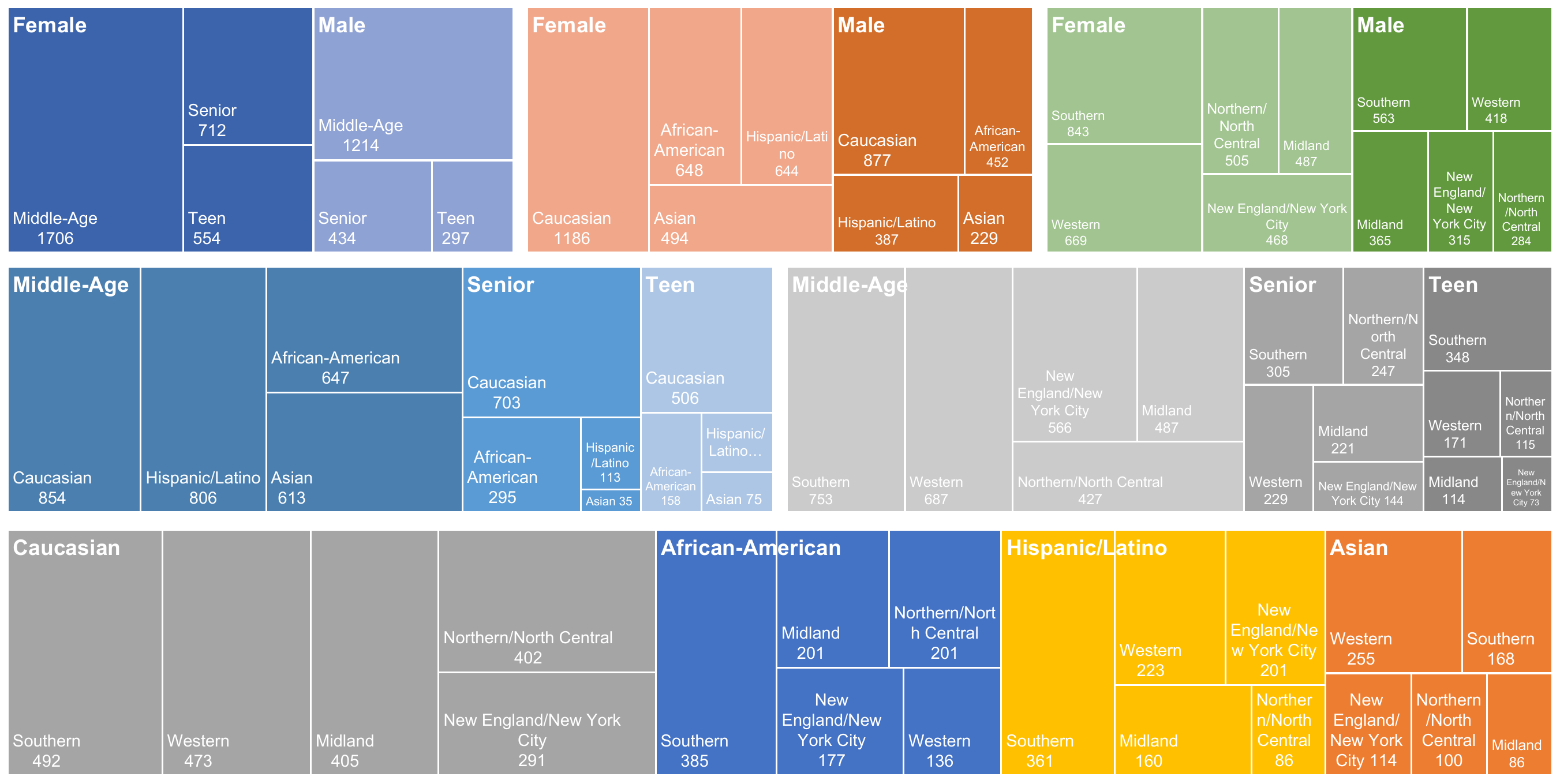}
    \centering
    \captionof{figure}{Data case distributions of the demographic subgroups in the FairLENS solo speaker transcription dataset.}
\label{apdx_fig1}
\end{minipage}}
\end{figure*}

\begin{figure*}[p]
\centering
\rotatebox{270}{\begin{minipage}{\textheight}
\includegraphics[width=\textwidth]{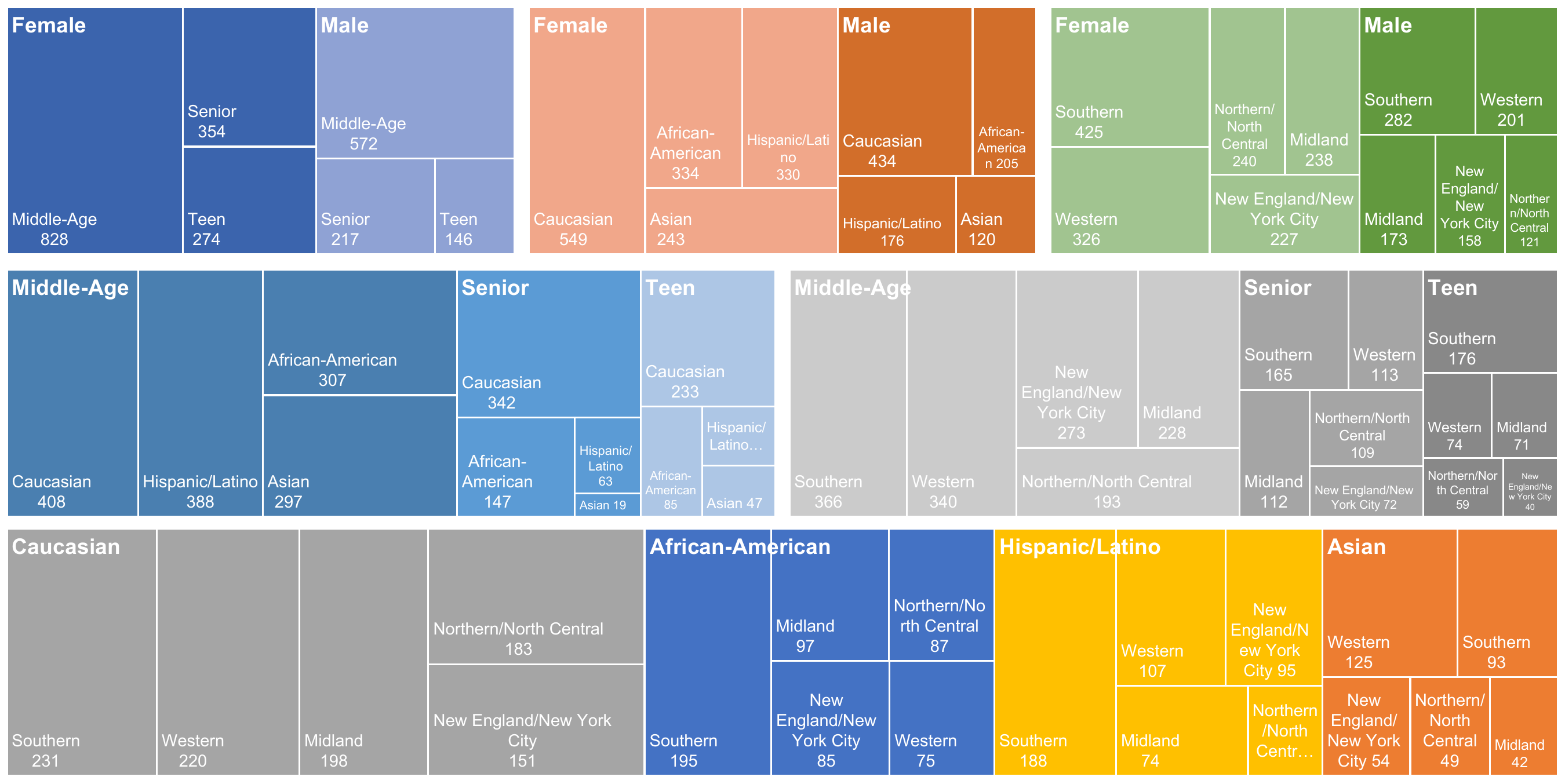}
    \centering
    \captionof{figure}{Data case distributions of the demographic subgroups in the FairLENS solo speaker dictation dataset.}
\label{apdx_fig2}
\end{minipage}}
\end{figure*}

\subsection{Dialogue Example}

Officer: I’m a police officer with the Boise Police Department. Do you wanna sit down or anything?

Victim: You can sit here. I'm going to grab his phone. I got his phone.

Officer: Okay, absolutely. Do you need to be checked out by any medics or anything?

Victim: No, I'm fine.

Officer: No?

Victim: I'm good.

Officer: So what happened to you?

Victim: So I was taking a delivery to the Brownville apartments.

Officer: Over here?

Victim: Yeah, it's on 20th and Brown. It's right on the corner there. I pull up to the loading zone right by the front of the door. There's these seats to the side of the door. There's a guy who has glasses and hair sitting right here, and there's a bald guy, sitting next to him.  I pull up and the bald guy immediately walks up to my car. I think he's like the guy who is getting the pizza. So I open the door, just a little bit, and I'm like, "Are you-" whatever the guy's name was, and he kind of said "no" but I didn't really understand him because he was obviously drunk, because he was just swaying at this point. And then he grabbed my door and opens it for me. I'm thinking, "Okay." I just grabbed  he pizza sitting next to me on the seat, and he's like reaching his hand out to take it, but I just kind of like walk past him. Then the guy who was sitting next to him stands up, and I'm like, "Oh, are you Pete?" And he's like, "Yeah, the pizza's for me." So I hand him the receipts because they have to sign their receipt. So I hand him the receipts and a pen, and as soon as I hand them, the drunk guy is standing right here and he pushes me on my shoulder and then just punches me in the face.

Officer: Yeah, you got some redness on your cheek.

Victim: Yeah, it feels a little puffy, but... It's fine. He just walks off. As he's walking off, I'm like, "I'm calling the police. Like, what the eff?" You know? And then I called it. And the guy who was with me said that he just sat down next to him and then threw his phone. Just threw it on the ground.

Officer: Okay.

Victim: And then he just started walking off towards 30th.

Officer: Would you recognize him if you saw him again?

Victim: If I saw him again, I definitely would.

Officer: Okay.
Victim: He wasn't walking very fast or anything. It didn't seem like he was trying to run away. He was... I don't know.

Officer: He started walking?

Victim: Yeah, he started walking.

Officer: He just hit you one time?

Victim: Just one time.

Officer: Do you know if it was closed fist?

Victim: Yeah, it was a closed fist.

Officer: Do you know which hand he hit you with?

Victim: No.

Officer: Are you pretty confident that they weren't together?

Victim: They definitely were not together. The guy said... I have his phone number in case you needed to talk to him, because he saw everything.

Officer: Okay.

Victim: Yeah, he waited with me and was saying, "Yeah, take his phone and give it to the police, because..." Yep.

Officer: Okay. Do you have a description of him?

Victim: Yeah, he's maybe like a little bit taller, like not a lot taller, and I'm 5'4", so maybe like 5'5", 5'6". And he's a white guy, maybe a little bit chunkier, like a weird build. And then... (laughs) he had these black glasses, like not circle but not square, and he had the thing around his glasses, and he was just bald.

Officer: Do you know what he was wearing at all?

Victim: Yeah, it was a long sleeve shirt, with these little indentations, not like a pattern but like a texture. It was either gray or like a green kind of shirt. And he was just wearing blue jeans. The other guy would probably be able to identify him, too, because he was sitting next to him while he was waiting for me.

Officer: And, how old do you think he is?

Victim: About 40?

\subsection{Freeform Prompt Instruction Example}

After a car accident, a driver is explaining the accident to an Officer. 

Option a) For one minute, describe an automobile accident that you were in, or were almost in, as if you were a driver explaining the accident to an Officer writing an incident report.

Option b) Think of a video of a car accident that you’ve seen or create an imaginary car accident situation in your mind. For one minute, describe the car accident as if you were a witness explaining the accident to an Officer. 

For either option, try to include:
\begin{itemize}
  \item Street names (real or fake)
  \item Speed limits (real or fake)
  \item Vehicle makes and models
  \item Vehicle damage and/or property damage and/or injuries
\end{itemize}

\begin{figure*}[!t]
    \begin{subfigure}{0.45\textwidth}
    \centering
    \includegraphics[width=\textwidth]{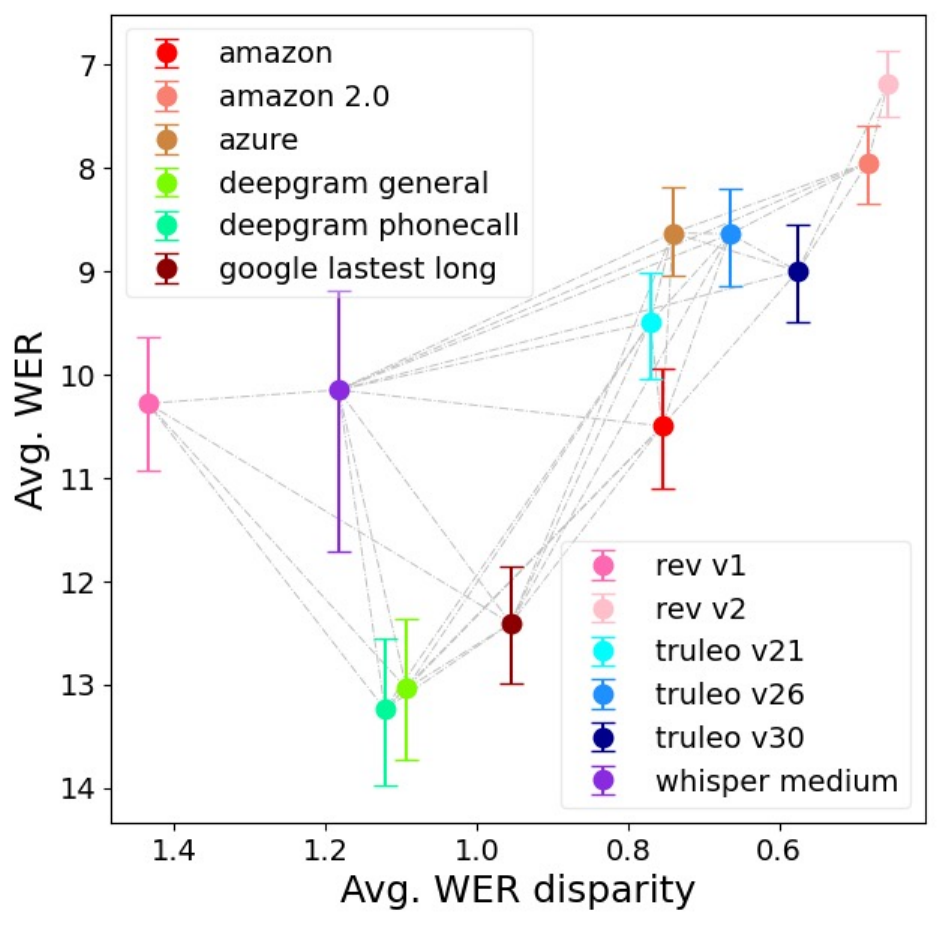}
    \caption{Indoor dialogue transcription dataset}
    \label{apdx_fig_3a}
    \end{subfigure}
    \hfill
    \begin{subfigure}{0.45\textwidth}
    \centering
    \includegraphics[width=\textwidth]{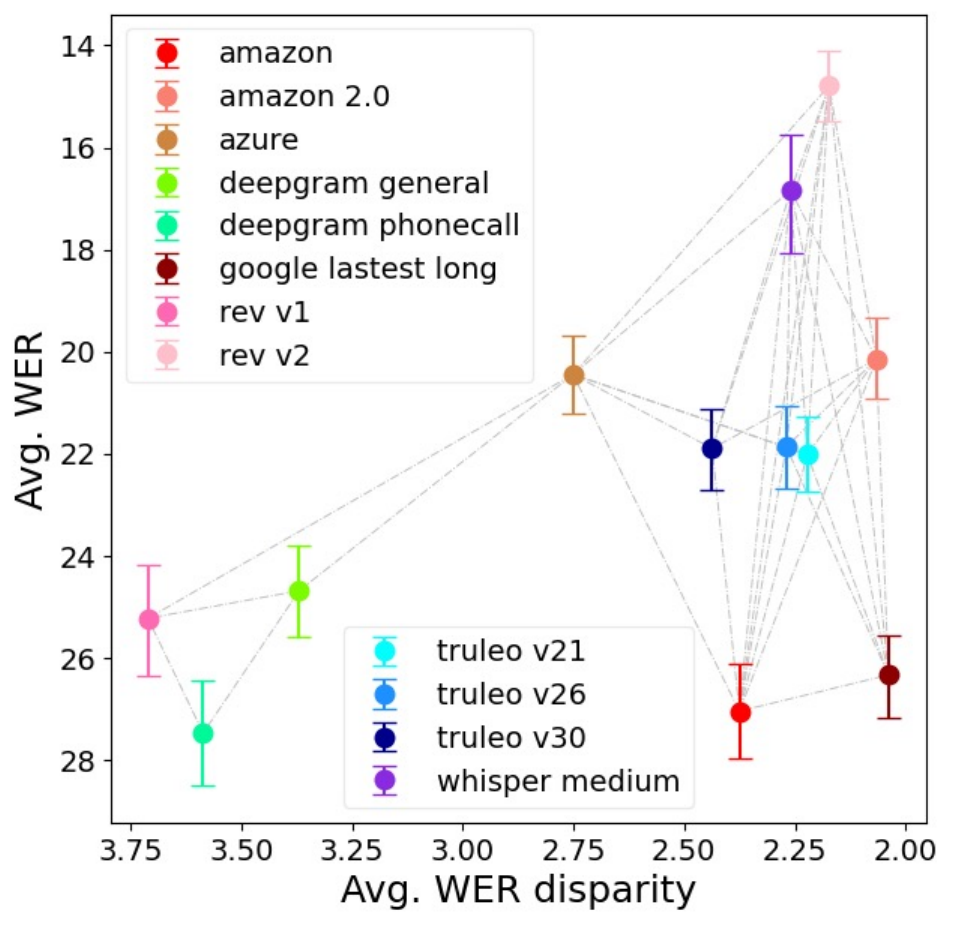}
    \caption{Outdoor dialogue transcription dataset}
    \label{apdx_fig_3b}
    \end{subfigure}
    \caption{Joint performance-fairness evaluation results.}
    \label{apdx_fig3}
\end{figure*}

\begin{figure*}[!t]
    \centering
    \includegraphics[width=\textwidth]{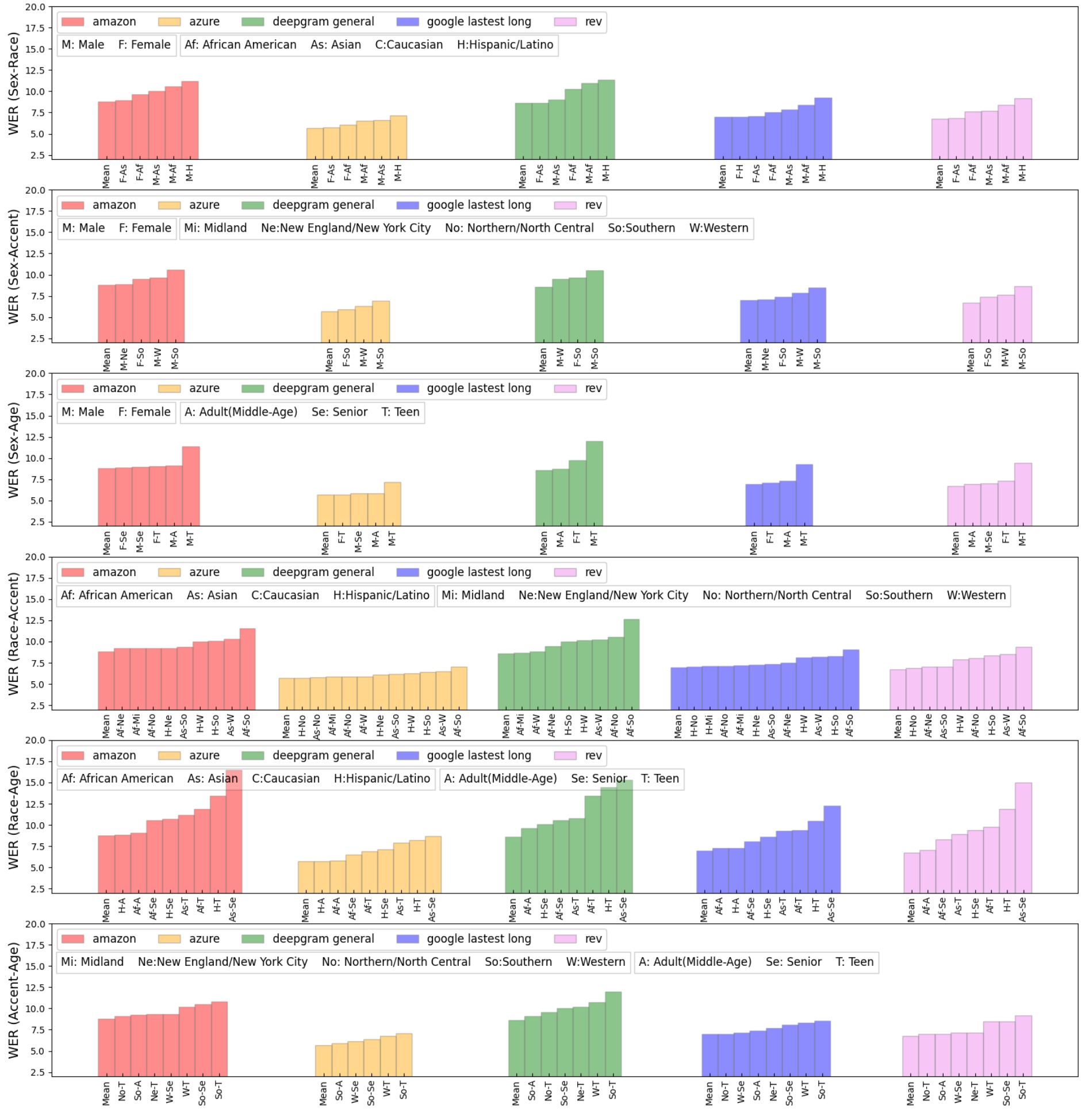}
    \caption{The performance of ASR models across different demographic subgroups on the solo speaker dictation dataset.}
    \label{apdx_fig4}
\end{figure*}

\subsection{Dictation Example}

This is Deputy Marcus period I responded to a noise complaint call at thirteen thirty on ten nine twenty one period When I arrived at the intersection of East Mayo Boulevard and North Tatum Boulevard in Phoenix comma I observed four juveniles standing on the corner period The four juveniles were holding various musical instruments period new line

They appeared to be talking to the driver of a silver Chrysler three hundred Arizona plate two four Young Ocean Union period As I pulled up to the boys comma the vehicle pulled away and went south on North Tatum Boulevard period new line

When questioned comma the juveniles stated that they had just been talking to their instructor who had told them they sounded like open quotes screeching cats close quotes and then had sent them outside to practice period new line

I advised the juveniles to move back inside period No further action taken period

\section{Data Distribution of Subgroups} 
Appendix Figure \ref{apdx_fig1} and \ref{apdx_fig2} shows the data distribution of different demographic subgroups.

\section{Dataset Comparison Settings}

\textbf{TIMIT dataset}: The accent groups categories are same with those in our FairLENS dataset. We rearrange race labels to three groups include White, Black, and Others due to extreme sparsity of data for other race groups. The sex groups in TIMIT is categorized to Female and Male.

\noindent \textbf{Casual Conversations datasets} The age attribute in the two Casual Conversations datasets are categorized to three groups: Teen ($\leq 20$), Adult($21-55$), and Senior ($\geq55$). We set a different cut-off age for the Teen and Adult since there are almost no participants of the Casual Conversations datasets that are under $18$. For the gender groups, only the data labeled as male or female of the gender groups are counted also because of the sparsity of the non-binary genders. The skin type groups follow the original setting in the datasets.

\section{ASR Models}
\textbf{Models tested for both Transcription (batch) \& Dictation (streaming)}
\begin{itemize}
  \item Amazon: Mid-2022 model version - \url{https://docs.aws.amazon.com/transcribe/latest/dg/what-is.html}
  \item Azure: Model Name: "20221013", Model ID: 5cb2efaa-4c4f-43f8-b7c4-cdadcc6ff8a9 - \url{https://learn.microsoft.com/en-us/azure/cognitive-services/speech-service/index-speech-to-text}
  \item Deepgram: General model - \url{https://deepgram.com}
  \item Deepgram: Phonecall model  - \url{https://deepgram.com}
  \item Google: Latest Long Model (multiple Google models were used, this was the best performing Google model for this use case) - \url{https://cloud.google.com/speech-to-text/docs}
  \item Rev "V2": Model version as of mid-2022 - \url{https://www.rev.com/services/speech-to-text-apis}
\end{itemize}

\textbf{Models tested only for Transcription (batch)}
\begin{itemize}
  \item Amazon "2.0": Late-2022 model version, model was not available for streaming at the time  - \url{https://docs.aws.amazon.com/transcribe/latest/dg/what-is.html}
  \item Rev "V1": 2021 model version - \url{https://www.rev.com/services/speech-to-text-apis}
  \item Truleo "v21": early-2022 model verison - \url{https://www.truleo.co}
  \item Truleo "v26": mid-2022 model verison - \url{https://www.truleo.co}
  \item Truleo "v30": late-2022 model verison - \url{https://www.truleo.co}
  \item OpenAI Whisper: Medium (769M param) model - \url{https://github.com/openai/whisper}
\end{itemize}

\section{Joint Performance-Fairness Evaluation Resuls on the Dialogue Datasets}

The performance and fairness results of the models on the dialogue datasets are shown in Figure \ref{apdx_fig3}. There is no obvious fairness cluster since we only conduct the experiments on the $12$ demographic groups (not on the $71$ subgroups). However, the Rev v2 model is the best one with the significantly lowest average WER on both datasets.

Another conclusion that can be drawn from Figure \ref{apdx_fig3} is that acoustic environment considerably influences both the performance and fairness of the models. The performance and fairness of all models have been degraded to varying degrees after switching from indoors to outdoors. In particular, the fairness shifting suggests that poor acoustic background further amplifies some models' biases. 

\section{Biases toward Specific Demographic Groups on the Dictation Dataset}
The performance of different ASR models across subgroups on the solo speaker dictation dataset are in Figure \ref{apdx_fig4}. The biases of ASR models towards specific groups are also observed, e.g., Asian, African American, Teens, and Southern Accent. 

\bibliographystyle{named}
\bibliography{ijcai24}

\end{document}